\documentclass[aps,prl,reprint,longbibliography,superscriptaddress, english]{revtex4-1}
 \usepackage[utf8]{inputenc}
 \usepackage{amsmath,bm}
 \usepackage{mathrsfs}
 \usepackage{amsfonts}
 \usepackage{graphicx}
 \usepackage{setspace} 
 \usepackage{graphicx}
 \usepackage{epstopdf}
 \usepackage{dcolumn}
 \usepackage{amsmath}
 \usepackage{epsfig}
 \usepackage{indentfirst}
 \usepackage{psfrag}
 \usepackage[normalem]{ulem}
 \usepackage{subfigure}
 \usepackage{amssymb}
\usepackage{caption}
\usepackage{color}
 \usepackage{units} 
\usepackage{soul} 
\usepackage{siunitx}
 \usepackage{graphicx}
 \usepackage{float}
 \graphicspath{{Figures/}} 
 \usepackage{dcolumn}
 \usepackage{bm}
 \usepackage{multirow}

\usepackage{physics}
\usepackage{dcolumn}
\usepackage{bm}
\usepackage{natbib}
\usepackage[backref=none,bookmarksnumbered=true,bookmarks=true,bookmarksopen=true,colorlinks=true,
citecolor=blue,linkcolor=blue,anchorcolor=green,urlcolor=blue,unicode=false]{hyperref}
\usepackage{verbatim}

\usepackage{ulem}[normalem] 

\def\BappCl{BaCl$_2$}

\newcommand{\str}[1]{\textcolor{cyan}{\st{#1}}}

\makeatletter
\newcommand\colorsout[1]{\bgroup \markoverwith{\textcolor{#1}{\rule[0.5ex]{2pt}{0.4pt}}}\ULon}

\makeatother

\newcommand{\bbonu}{\ensuremath{\beta\beta0\nu}}
\newcommand{\bbtnu}{\ensuremath{\beta\beta2\nu}}

\newcommand{\XE}{\ensuremath{{}^{136}\rm Xe}}
\newcommand{\GE}{\ensuremath{{}^{76}\rm Ge}}

\newcommand{\TE}{\ensuremath{{}^{128}\rm Te}}

\newcommand{\Bapp}{\ensuremath{{\rm Ba^{2+}}}}

\newcommand{\Nap}{\ensuremath{\rm Na^{+}}}

\DeclareSIUnit\c{\mbox{$c$}}
\DeclareSIUnit\magn{\mbox{$\times$}}
\DeclareSIUnit\min{min}
\DeclareSIUnit\week{week}
\DeclareSIUnit\year{yr}
\DeclareSIUnit\years{years}
\DeclareSIUnit\yr{yr}
\DeclareSIUnit\standard{std}
\DeclareSIUnit\str{sr}
\DeclareSIUnit\ppm{ppm}
\DeclareSIUnit\ppb{ppb}
\DeclareSIUnit\ppt{ppt}
\DeclareSIUnit\pe{PE}
\DeclareSIUnit\spe{SPE}
\DeclareSIUnit\ev{events}
\DeclareSIUnit\ct{counts}
\DeclareSIUnit\neutron{\mbox{$n$}}
\DeclareSIUnit\smp{samples}
\DeclareSIUnit\Sample{S}
\DeclareSIUnit\ch{ch}
\DeclareSIUnit\hit{hit}
\DeclareSIUnit\hits{hits}
\DeclareSIUnit\bin{(\mbox{5-PE}~bin)}
\DeclareSIUnit\sgm{\mbox{$\sigma$}}
\DeclareSIUnit\rms{RMS}
\DeclareSIUnit\keVr{\mbox{keV$_{\rm nr}$}}
\DeclareSIUnit\keVee{\mbox{keV$_{e{\rm e}}$}}
\DeclareSIUnit\ph{photon}
\DeclareSIUnit\pes{pes}
\DeclareSIUnit\el{electrons}
\DeclareSIUnit\pm{PMT}
\DeclareSIUnit\inch{"}
\DeclareSIUnit\bit{bit}
\DeclareSIUnit\sample{samples}
\DeclareSIUnit\barn{barn}
\DeclareSIUnit\bara{bar}
\DeclareSIUnit\barg{barg}
\DeclareSIUnit\mlardepth{\mbox(meter~of~\LAr~depth)}
\DeclareSIUnit\Curie{Ci}
\DeclareSIUnit\psi{psi}
\DeclareSIUnit\parsec{pc}
\DeclareSIUnit\liveday{\mbox{live-days}}
\DeclareSIUnit\days{\mbox{days}}
\DeclareSIUnit\day{\mbox{day}}
\DeclareSIUnit\miles{\mbox{miles}}
\DeclareSIUnit\degreeC{\mbox{$^{\circ}$C}}
\DeclareSIUnit\electron{\mbox{$e^-$}}
\DeclareSIUnit\Euro{\mbox{\euro}}
\DeclareSIUnit\cph{cph}
\DeclareSIUnit\neq{neq}
\DeclareSIUnit\unit{unit}
\DeclareSIUnit\byte{Byte}
\DeclareSIUnit\Bq{\becquerel}

\begin{document}
\title{ \Bapp\ ion trapping by organic submonolayer: towards an ultra-low background neutrinoless double beta decay detector}

\author{P.~Herrero-G\'omez}
\affiliation{Centro de F\'isica de Materiales (CSIC-UPV/EHU), San Sebasti\'an / Donostia, E-20018, Spain}
\affiliation{Donostia International Physics Center DIPC, San Sebasti\'an / Donostia, E-20018, Spain}
\author{J.P.~Calupitan}
\affiliation{Centro de F\'isica de Materiales (CSIC-UPV/EHU), San Sebasti\'an / Donostia, E-20018, Spain}
\author{M.~Ilyn}
\affiliation{Centro de F\'isica de Materiales (CSIC-UPV/EHU), San Sebasti\'an / Donostia, E-20018, Spain}
\author{A.~Berdonces-Layunta}
\affiliation{Centro de F\'isica de Materiales (CSIC-UPV/EHU), San Sebasti\'an / Donostia, E-20018, Spain}
\affiliation{Donostia International Physics Center DIPC, San Sebasti\'an / Donostia, E-20018, Spain}
\author{T.~Wang}
\affiliation{Centro de F\'isica de Materiales (CSIC-UPV/EHU), San Sebasti\'an / Donostia, E-20018, Spain}
\affiliation{Donostia International Physics Center DIPC, San Sebasti\'an / Donostia, E-20018, Spain}
\author{D.~G. de Oteyza}
\affiliation{Centro de F\'isica de Materiales (CSIC-UPV/EHU), San Sebasti\'an / Donostia, E-20018, Spain}
\affiliation{Donostia International Physics Center DIPC, San Sebasti\'an / Donostia, E-20018, Spain}
\author{M.~Corso}
\affiliation{Centro de F\'isica de Materiales (CSIC-UPV/EHU), San Sebasti\'an / Donostia, E-20018, Spain}
\affiliation{Donostia International Physics Center DIPC, San Sebasti\'an / Donostia, E-20018, Spain}
\author{R.~Gonz\'alez-Moreno}
\affiliation{Donostia International Physics Center DIPC, San Sebasti\'an / Donostia, E-20018, Spain}
\author{I.~Rivilla}
\affiliation{Donostia International Physics Center DIPC, San Sebasti\'an / Donostia, E-20018, Spain}
\affiliation{Ikerbasque, Basque Foundation for Science, Bilbao, E-48009, Spain}
\author{B.~Aparicio}
\affiliation{Department of Organic Chemistry I, University of the Basque Country (UPV/EHU), Centro de Innovaci\'on en Qu\'imica Avanzada (ORFEO-CINQA), San Sebasti\'an / Donostia, E-20018, Spain}
% ...


\begin{abstract}

If neutrinos are their own antiparticles \cite{Majorana:1937}, the otherwise-forbidden nuclear reaction known as neutrinoless double beta decay (\bbonu) can occur, with a characteristic lifetime which is expected to be very long, making the suppression of backgrounds a daunting task. It has been shown that detecting (``tagging'') the \Bapp\ dication produced in the double beta decay ${}^{136}\mathrm{Xe} \rightarrow {}^{136}\Bapp + 2 e + (2 \nu)$ in a high pressure gas experiment, could lead to a virtually background free experiment \cite{Nygren_2015,Jones:2016qiq, McDonald:2017izm, rivilla_fluorescent_2020}. To identify these \Bapp, chemical sensors are being explored as a key tool by the NEXT collaboration \cite{Thapa:2019zjk, rivilla_fluorescent_2020,thapa_demonstration_2021}. Although used in many fields, the application of such chemosensors to the field of particle physics is totally novel and requires experimental demonstration of their suitability in the ultra-dry environment of a xenon gas chamber. Here we use a combination of complementary surface science techniques to unambiguously show that \Bapp\ ions can be trapped (chelated) in vacuum by an organic molecule, the so-called fluorescent bicolour indicator (FBI) \cite{rivilla_fluorescent_2020} (one of the chemosensors developed by NEXT), immobilized on a surface. We unravel the ion capture mechanism once the molecules are immobilised on Au(111) surface and explain the origin of the emission fluorescence shift associated to the trapping of different ions. Moreover, we prove that chelation also takes place on a technologically relevant substrate, as such, demonstrating the feasibility of using FBI indicators as building blocks of a \Bapp\ detector.

\end{abstract}
\date{\today}
\maketitle

The rare neutrinoless double beta  (\bbonu) decay, $(Z,A) \rightarrow (Z+2,A) + 2\ e^{-}$, can occur if and only neutrinos are Majorana particles \cite{Majorana:1937}, i.e., identical to their antiparticles. An unambiguous observation of such a decay would have deep implications in particle physics and cosmology \cite{Sakharov1967,Fukugita:1986hr,GellMann:1980vs,Yanagida:1979as,Mohapatra:1979ia}. 
The conventional double beta decay (\bbtnu), in which two neutrinos are emitted in addition to the electrons, occurs in a handful of isotopes, some of which also offer the necessary features (a reasonable isotopic abundance, a decay energy sufficiently high, etc.) to be used as sources/targets in experiments seeking to observe the \bbonu\ decay. Examples of such isotopes which have been used for large-scale searches are \GE, \TE, and \XE. All attempts to detect a signal so far have not succeeded, and experimental bounds in excess of $10^{26}$ years have been set for the most sensitive searches based on \XE\ and \GE\ \cite{Gando:2016ji, Agostini:2018tnm}.
The field is currently aiming to increase the sensitivity by at least one, and eventually two or more orders of magnitude \cite{Gomez-Cadenas:2019sfa}. This, in turn, implies large exposures, measured in ton-years, and even more importantly, a greatly enhanced capability to suppress radioactive backgrounds. 

In a high pressure xenon gas time projection chamber (HPXe-TPC) such as those being developed by the NEXT experiment \cite{Martin-Albo:2015rhw, NEXT:2020amj}, the double beta decay of \XE\ will create a $^{136}$\Bapp\ dication and two electrons.
The decay mode with two neutrinos, \bbtnu, has been observed in xenon with a lifetime of the order of $2 \times 10^{21}$ years \cite{Ackerman:2011gz}. The  signal is the same as that of \bbonu, except for the total energy of the electrons, which is a continuous distribution for the \bbtnu\ case and spikes around the decay energy, Q$_{\beta\beta}$ (about 2.45 MeV in the case of \XE) for \bbonu\  \cite{rivilla_fluorescent_2020}. The excellent energy resolution of NEXT allows suppressing the contamination of \bbtnu\ to \bbonu\ by at least nine orders of magnitude \cite{ALVAREZ2013101,Renner:2019pfe,rivilla_fluorescent_2020}, and thus, \bbtnu\ is not a significant background for NEXT up to lifetimes of \bbonu\ in the order of $10^{30}$ years. 

In the TPC, the electric field that drifts the ionization electrons from the two emitted electrons towards the detector anode, will cause the \Bapp\ dication to drift towards the cathode \cite{Bainglass:2018odn}. Thus, it is feasible to trigger on interesting events (e.g, those with sufficiently high energy). The barycenter of the electron tracks can be used to predict the impact position of the ion, or alternatively, ion transport devices, known as RF carpets could be used to direct the ion to a specific region in the cathode \cite{NEXT:2021idl}. Furthermore, it is possible to correlate the arrival time of the dication to the cathode with the electrons detected in the anode. Such coincidence could lead to a virtually background free experiment.

Although the barium tagging concept was proposed more than three decades ago \cite{Moe:1991ik}, a practical way to detect \Bapp\ {\it in situ} in a HPXe-TPC was only conceived recently \cite{Nygren_2015, Jones:2016qiq}. The idea relies on the capability of fluorescent molecules of changing their optical properties upon detecting target analytes \cite{valeur_chemical,wolfbeis_materials_2005}. An initial proof of concept \cite{McDonald:2017izm} resolved individual \Bapp\ dications in aqueous solution using Fluo-3, a well known commercial indicator. A suitable barium detector in NEXT requires a functionalised surface that must include a monolayer of the molecular sensor and must efficiently operate in a noble gas atmosphere. Over the last three years an intense R\&D program to develop chemosensors able to form a supramolecular complex with \Bapp\ in dry medium has been carried out \cite{Thapa:2019zjk, rivilla_fluorescent_2020,thapa_demonstration_2021}. To date, no experiments have been conducted in which the processes of chelation and detection occur fully under such conditions.

One of the sensors developed by NEXT, the so-called
Fluorescent Bicolor Indicator (FBI) \cite{rivilla_fluorescent_2020} combines an enhanced fluorescence with a shift of the emission spectrum (about 61 nm towards the blue) when the indicator is complexed with \Bapp.  This is due to the specific molecular design of the fluorescent indicator having a crown ether connected to a benzo[\textit{a}]imidazo[5,1,2-\textit{cd}]indolizine fluorophore by a phenyl group. The benzoimidazoindolizine group has been shown to have highly tunable bright emission \cite{Stasyuk_benzo,Levesque_general} while the crown ether is capable of interacting with \Bapp\ ions \cite{valeur_chemical,maleknia_cavity-size-dependent_2002}. In the presence of a \Bapp\ ion, calculations predict that coordination happens between the cation, one nitrogen atom of the fluorophore, the crown ether and the para-substituted phenyl ring, causing a  {torsion-induced decoupling between non-coplanar components of the fluorophore with respect} to the non-chelated molecule. This induces a large change in the electronic properties of the dye, causing a blue-shift in emission, which can be used to filter the signal of chelated indicators.

\begin{figure*}[ht!]
	\includegraphics[width=1\textwidth]{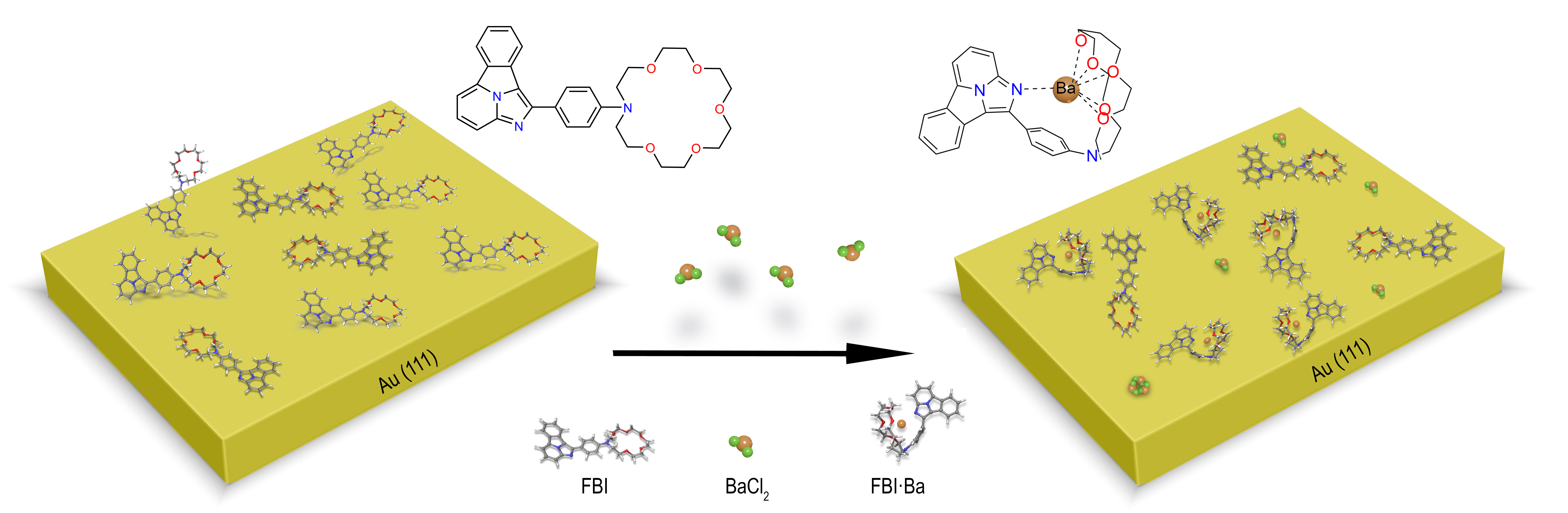}
	\caption{\label{ModeloFBI} 
    Model of the FBI molecules before and after chelation and schematic representation of the experiment we have carried out: FBI molecules were sublimated on a Au(111) surface, chelated in-situ and characterized inside the UHV chamber.}
\end{figure*}

In this paper we combine two highly sensitive surface techniques: X-ray Photoemission Spectroscopy (XPS) and Scanning Tunnelling Microscopy and Spectroscopy (STM/STS), to prove how different ions interact with FBI molecules deposited on suitable substrates. We demonstrate that only \Bapp\ ions induce molecular structural changes, modifying the electronic structure in ways that would affect the fluorescence emission at suitable surfaces. Coordination with crown-ether happens entirely in ultra-high vacuum (UHV) (see model in Figure {\ref{ModeloFBI}}), which ensures that chemical, structural and electronic changes happen in the absence of solvents, air molecules or spurious contaminants. This is, therefore, a crucial step toward the development of a \Bapp\ detector. 

All the indicators developed by the NEXT experiment are based on crown ethers. Because of their capability to capture a variety of guest species, including metal cations, protonated species and neutral and ionic molecules \cite{dobler1981ionophores}, crown ethers \cite{gokel_crown_1991} have been extensively used to recognise and trap metal or molecular ions \cite{more_intrinsic_1999}, \cite{maleknia_cavity-size-dependent_2002}. However, they have been poorly studied in solid state. Few examples can be found in the literature where self-assembled monolayers of crown ether derivatives have been grown and used on surfaces. Moreover, in all previous studies, either the growth or ion trapping or both have taken place in solution \cite{yoshimoto_hostguest_2003}, \cite{flink_recognition_1999}, \cite{inokuchi_new_2015}. 
 There are only two works, as far as we know, where crown ethers were deposited under UHV conditions \cite{feng_growth_2018} and their metal trapping capability was proven also under UHV \cite{stredansky_-surface_2019}. Thus, in addition to the progress relevant for a future \bbonu\ experiment, the work presented here advances substantially the understanding of the physico-chemical properties of crown ethers immobilised on solid surfaces.

\section{Molecular sublimation in vacuum}

In order to characterize the ion trapping capability of the FBI molecules, first it is mandatory to deposit these FBI molecules on a surface by sublimation in UHV and characterize them structural and chemically. Thus, we chose a gold substrate, the Au(111) face (single crystal cut to present (111) close-packed planes parallel to the surface), because it is a well known noble metal surface where very low substrate-molecule interaction was expected. To determine their chemical composition, we used XPS. It is worth mentioning that, since XPS sensitivity is limited to a depth of a few nanometers, molecular coverage below 1 monolayer (ML) were always used in this study to ensure access to the substrate core levels for calibration. 

\begin{figure*}[ht!]
	\includegraphics[width=0.9\textwidth]{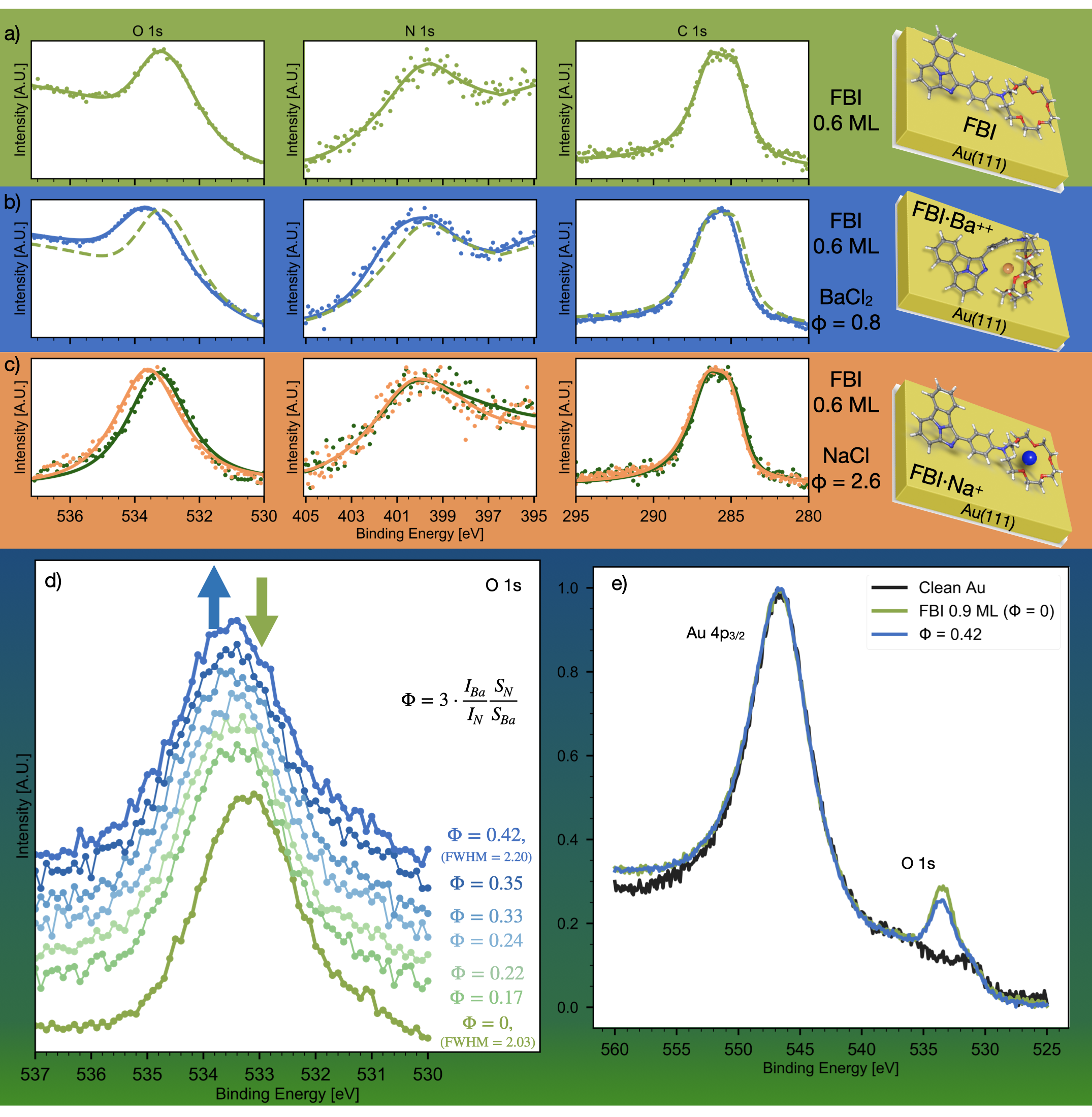}
	\caption{\label{XPS_FBI_Au} 
    XPS demonstration of the chemical changes induced by the molecular chelation: (a) O 1s, N 1s and C 1s core level spectra measured after sublimation of 0.6 ML of FBI deposition on Au(111); (b) O 1s, N 1s and C 1s core level spectra measured on the previous sample after chelation with 0.80 \Bapp ions per FBI molecule; (c) O 1s, N 1s and C 1s core level spectra measured after 0.6 ML of FBI deposited on Au(111) (green spectra) and after chelation with \Nap\ (2.60 \Nap ions per FBI molecule) (orange line). For the three panels, dots correspond to raw values and solid lines to fitted curves (the fitting procedure is discussed in the Methods section). (d) Evolution of O 1s core level as a function of the \Bapp deposition on 0.9 ML of FBI on Au(111). The spectra were manually shifted in the y-axis to better show the evolution. The O 1s core level region is displayed in (c) and (d) after subtraction of the contribution from Au 4p 3/2, in order to emphasize the spectral changes in the series; (e)  Au 4p 3/2 and O 1s shown previous to any data treatment.}
\end{figure*}

Figure {\ref{XPS_FBI_Au}}a shows the XPS spectra of the three molecular core levels of FBI, i.e. O 1s, N 1s and C 1s. The spectra were measured for 0.6 ML deposited on Au(111) at room temperature (RT). The C 1s core level can be fitted using two components, one {centered} at around 284.7 eV and a second and more intense one at 286.3 eV. The former component corresponds to C-C bonds whereas the component at higher binding energy (BE) includes contributions from C-O and C-N bonds, with their relative intensities in agreement with having intact molecules. In the N 1s region, around 400.4 eV, a faint peak is visible. The position of the maximum is compatible with the {enamine-imine} groups of the molecular composition. Finally, the O 1s core level presents a single component peak centered at 533.0 eV, which is compatible with previous reports on closely related crown ether groups \cite{stredansky_-surface_2019}. The ratios between the core levels, C/O = 6.2, C/N = 10.3, are in agreement with having molecules of $ \mathrm{C_{31}N_{3}O_{5}H_{35}}$ stoichiometry on the surface. 

\begin{figure*}[ht!]
	\includegraphics[width=0.99\textwidth]{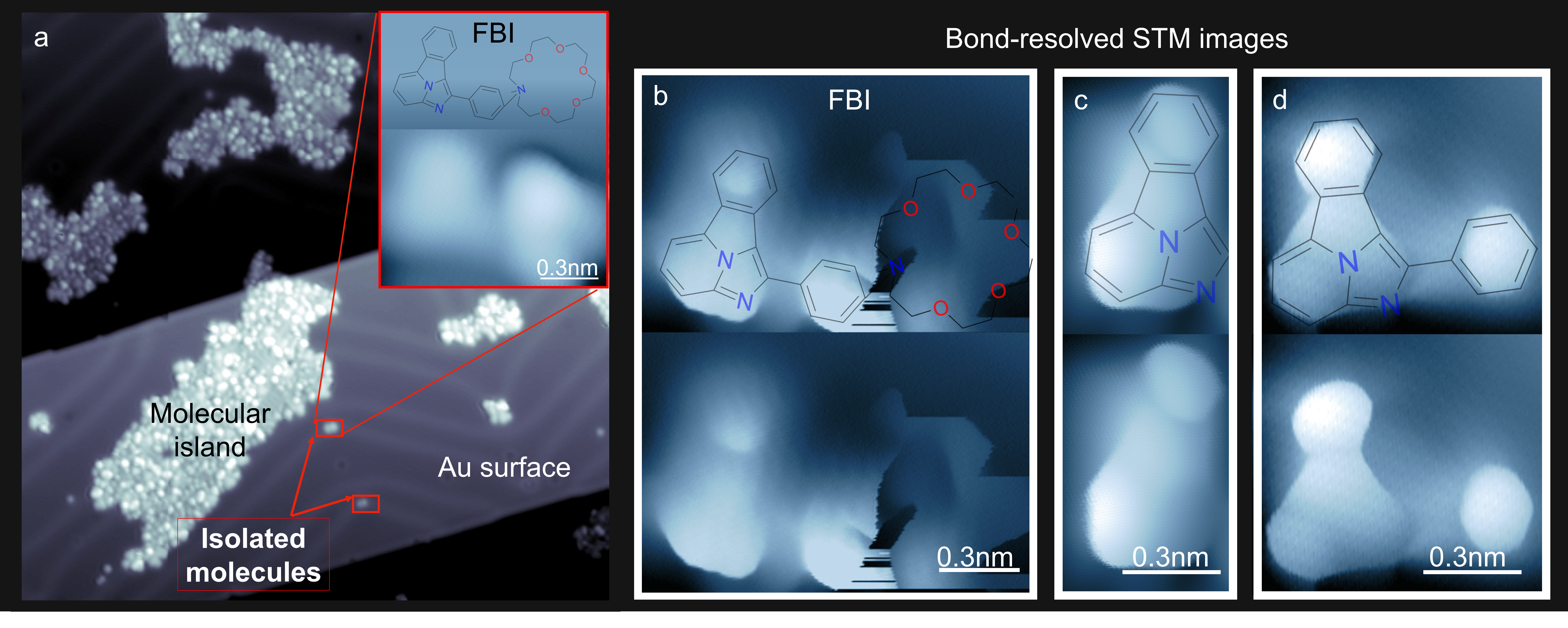}
	\caption{\label{FIG_BRSTM} 
    (a) Large scale STM image ($50\times50$ nm$^2$) of around 0.4 ML of FBI molecules deposited on Au(111) surface and measured at 4 K. Red squares show isolated molecules. Inset: constant current zoom STM image of a FBI molecule (I = 60 pA / U = 1.4 V) deposited on Au(111). (b-d) Bond-resolved STM image measured with a CO-functionalized probe (constant height, U = 5 mV) of individual FBI molecules: (b) benzo[\textit{a}]imidazo[5,1,2-\textit{cd}]indolizine fluorophore with phenyl ring and aza-crown ether (exactly the same molecule shown in the inset in (a)); (c)  benzo[\textit{a}]imidazo[5,1,2-\textit{cd}]indolizine fluorophore; (d) benzo[\textit{a}]imidazo[5,1,2-\textit{cd}]indolizine fluorophore with phenyl ring but without the crown ether (d). For clarity, the same image is shown twice, one with and one without the molecular model superimposed to guide the eyes.}
\end{figure*}

In addition to the XPS analysis, we have confirmed the intact sublimation of the molecules by STM. Deposition of a submonolayer of FBI resulted mostly in disordered islands, coexisting with isolated molecules (Figure {\ref{FIG_BRSTM}a}). STM images of the single molecules (inset in Figure  {\ref{FIG_BRSTM}a}) show two main triangular-like lobes. Based on the molecular structure, one of them corresponds to the fluorophore and the other to the crown-ether moieties bonded to the phenyl ring. The apparent height of both is different, with one of them appearing lower and flatter than the other. In order to correlate the apparent molecular shape and the molecular structure, bond-resolved STM images were taken. For that, the STM tip was functionalized with CO molecules operating in the repulsive tip-sample interaction regime, a method extensively used in STM field for characterizing planar organic molecules \cite{gross_recent_2011,gross_atomic_2018}. Application of this measurement mode to the FBI results in the images displayed in Figure \ref{FIG_BRSTM}b. On the right, the internal bonding structure of the fluorophore is approximately resolved, while the aza-crown ether regions is not well resolved. Due to its larger conformational flexibility and its non-planarity, the tip stability is not good enough and, as a consequence, the contrast observed in the aza-crown ether part is an artefact related to excessively strong interactions with the flexible CO tip-apex. \cite{moll_mechanisms_2010,hapala_mechanism_2014}. To confirm the imaging signature of the fluorophore, we also sublimated two FBI derivatives specifically synthesized without the aza-crown ether component. Figures \ref{FIG_BRSTM}c and d show the bond-resolving images of FBI derivatives sublimated on Au(111) surface: the fluorophore (Figure {\ref{FIG_BRSTM}c}) and the fluorophore with the phenyl ring but without the aza-crown ether (Figure {\ref{FIG_BRSTM}d}). The absence of the aza-crown ether allowed the molecules on the surface to adopt a planar, rigid conformation. This clearly improved the clarity of the images, with the carbon framework of the molecule well visualised. In turn, this supports our previous assignment of the different parts when measuring the original FBI molecules and confirms their unchanged chemical structure upon sublimation.  

\section{Chemical demonstration of chelation}

Once confirmed the presence of intact FBI molecule on the surface, \BappCl\ was sublimated to test the molecular chelation. Prior to the sublimation on the FBI, \BappCl\ was sublimated on clean Au(111) surface to confirm its stoichiometry. By comparing the core level intensities of Ba 3d 5/2 and Cl 2p core levels (taking into account the corresponding sensitivity factor of each element), the ratio Ba:Cl was 1:2 as expected. Surprisingly, when \BappCl\ was deposited on FBI-functionalised Au(111), this ratio was around 1:1.5$\pm{0.2}$, meaning that some chlorine atoms desorbed when they reached the sample. From simulations \cite{rivilla_fluorescent_2020}, the chloride ions are expected to behave as passive spectators in the chelation, which is consistent with their desorption when the FBI molecule trapped the \Bapp. Because of this lack of stoichiometry, we refrain from using \Bapp\ ML to refer to the amount of sublimated molecules and, instead, we will refer the \Bapp\ ions per FBI molecule. The \Bapp\ dose was estimated using the ratio between XPS intensities of the Ba 3d 5/2 and N 1s core levels divided by the corresponding sensitivity factor, and always taking into account that there are 3 N atoms per molecule.

After sublimation of 0.80 \Bapp\ ions per FBI molecule, the core levels slightly shift, which indicates changes in the chemical environment. Figure {\ref{XPS_FBI_Au}b} shows the O 1s, N 1s, and C 1s, whose shifts toward higher BE are distinguished, mainly on O 1s. The O 1s core level exhibits an upward chemical shift of 0.5 eV (0.4eV for N 1s). By curve fitting deconvolution (discussed in the Methods section) we determine that this shift is not just a doping shift, but a real chemical change, inferred by the growth of a new component. This is confirmed by the evolution of the O 1s for very low \Bapp\ doses on the surface. Figure {\ref{XPS_FBI_Au}d} shows the evolution of the O 1s for incremental amounts of \Bapp\ ions from 0 to almost 0.5 \Bapp\  ions per FBI molecule, i.e. up to about 2 FBI {molecules} per 1 \Bapp. Once \BappCl\ is added to the sample, in the O 1s core level a second component grows at higher binding energy, which induces an increase of the peak width. At the latest stage of barium addition, the FWHM increases by 9\% with respect to the initial state, which indicates that an extra component appears. We deconvoluted the peaks after \Bapp\ addition using two components, one at 532.95 eV (FWHM = 2.20 eV), corresponding to the FBI molecules and another at around 533.9 eV, consistent with the molecules undergoing a chemical change upon chelation, as previously observed for crown ether chelation with Na \cite{stredansky_-surface_2019}. It is important to mention the difficulty of quantifying correctly the variation of one component with respect to the other for O 1s. This is due to the close proximity of the Au 4p core level right tail, Figure {\ref{XPS_FBI_Au}d}, with a contribution in the O 1s region.

To evaluate whether the nature of the ions has any influence on the oxygen core level shift for the FBI molecules, we tested the chelation with \Nap. Thus, Figure {\ref{XPS_FBI_Au}c} shows the O 1s, N 1s and C 1s core levels measured on 0.6 ML of FBI on Au(111) before and after the sublimation of 2.60 \Nap\ per FBI. The maximum of the O 1s core level shifts again for \Nap-chelated FBI molecules 0.5 eV toward higher BE, confirming that O 1s shift can be considered as a fingerprint of the chelation. Moreover, there is no apparent shift on the N 1s (nor C 1s) core level, which reveals that N 1s is not playing any role in the chelation process. According to theoretical calculations, upon chelation with \Bapp, FBI undergoes a structural torsion generated by the interaction between the ion, the iminic N atom of the benzo[\textit{a}]imidazo[5,1,2-\textit{cd}]indolizine moiety and the phenyl group of the unbound fluorophore \cite{rivilla_fluorescent_2020}. On the contrary, in the case of \Nap\ calculations do not show such an important molecular distortion because of the smaller size of the ion. Thus, this difference between \Bapp\- and \Nap-chelation can be interpreted as a first proof that the molecular conformation of FBI and the cation-phenyl interaction varies depending on the nature of the trapped ions. 

\section{Molecular structural rearrangement induced by chelation}

\begin{figure*}[ht!]
	\includegraphics[width=0.9\textwidth]{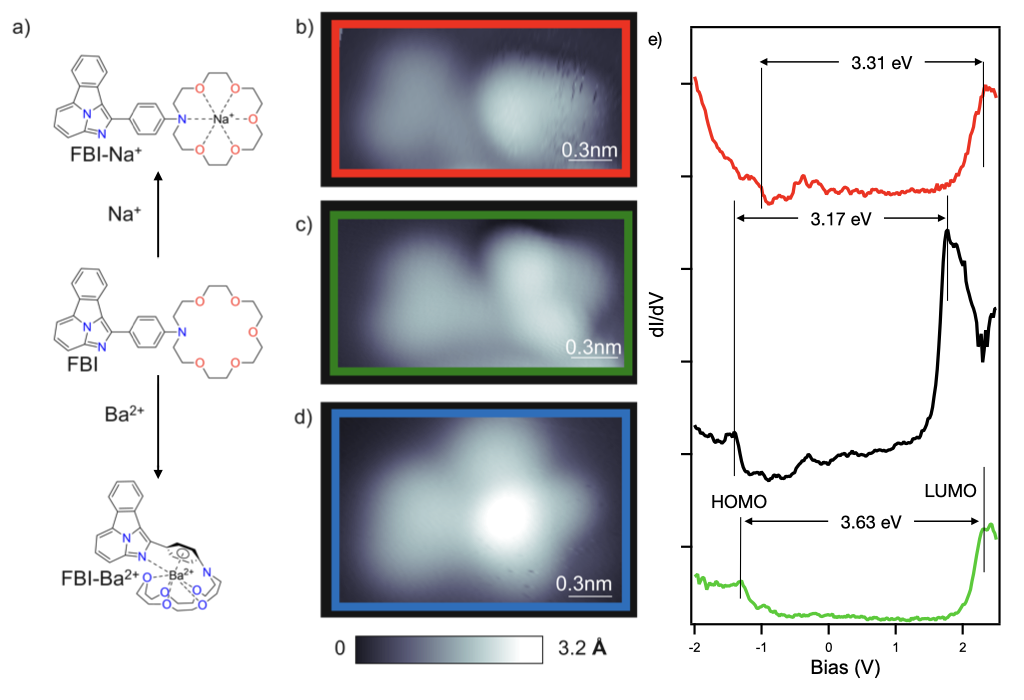}
	\caption{\label{Fig_STS} 
    (a) Schematic diagram of metal intercalation. (b-d) STM images of (b) \Nap-complexed (U = 0.9 V, I = 60 pA), (c) native (U = 1.4 V, I = 60 pA), and (d) \Bapp-complexed (U = -1.8 V, I = 20 pA) FBI. (Scale bars = 0.3 nm) (e) STS spectra of native (green), \Nap-complexed (red), and \Bapp-complexed (blue) FBI.}
\end{figure*}

In order to visualize the aforementioned structural changes undergone by the FBI molecules upon chelation, 
 STM experiments were, again, carried out. Figure \ref{Fig_STS}a shows the schematic representation of the FBI molecules before and after chelation with \Nap\ and \Bapp, and Figure \ref{Fig_STS}b-d displays representative STM images next to each of the cases. In all of them, the fluorophore and the aza-crown ether region are distinguished. The \Nap-chelated FBI image remains almost unchanged compared to the non-chelated FBI, while the FBI chelated with \Bapp\ has a more complex shape. Although the interpretation of constant current mode STM images is always difficult to obtain because of the contrast determined by a convolution of topographical and electronic contributions, here we can directly relate the images differences with the expected conformational changes, since the images were measured at bias voltages well inside the molecular gap, i.e., with no variation in the electronic structure. Taking into account that the three STM images are plotted with a common colour code, it can be directly observed that the apparent heights of the crown ether follow the expected trend (higher for \Bapp\ chelated molecule). Indeed, the apparent height for each molecule, as measured on the highest point of the crown, is 2.5$\pm$ 0.1 \AA, 2.1$\pm$ 0.1 \AA, and 2.9$\pm$ 0.1 \AA, for the pristine FBI and the \Nap- and \Bapp-chelated molecules respectively. Thus, complexation of crown ethers with alkali metals like \Nap\ causes the oxygen atoms to point to the center, forcing the ring to adopt a flatter conformation relative to the native molecule. Instead, FBI molecule adopts a more three-dimensional conformation upon chelation with \Bapp.  
 
\begin{table}[]
    \centering
    \begin{tabular}{|c|c|c|c|c|}
        \hline
        Species &  Band gap / eV (nm) & $\lambda^{\mathrm{abs}}_{\mathrm{max}}$ / \text{nm} & $\lambda^{\mathrm{emi}}_{\mathrm{max}}$/\text{nm} \\ \hline
        FBI & 3.17 (391.1) & 432.5 & 489 \\
        FBI-\Nap & 3.31 (374.6) & 430 & 489 \\
        FBI-\Bapp & 3.63 (341.6) & 420.5 & 428 \\ \hline
    \end{tabular}
    \caption{Band gap values measured by STS on Au(111) vs. absorption ($\lambda^{\mathrm{abs}}_{\mathrm{max}}$) and fluorescence emission ($\lambda^{\mathrm{emi}}_{\mathrm{max}}$) spectral peaks measured in solution.}
    \label{tab:bandgaps}
\end{table}

As we mentioned before, the distortion of the molecule upon metal coordination has an important impact on the fluorescence response. The lowest fluorescence emission peak of the non-chelated FBI molecule comes from the de-excitation of electrons from the Lowest Unoccupied Molecular Orbital (LUMO) to the Highest Occupied Molecular Orbital (HOMO). The torsion of the phenyl group decreases the effective conjugation, thus increases the symmetry allowed $\pi \rightarrow \pi^*$ gap, resulting in an increase of the HOMO-LUMO gap, and therefore introducing a blue shift of the fluorescent emission. In order to check that, we use scanning tunneling spectroscopy (STS) to scan the local density of states of the system. In this way, it is possible to detect the changes in the HOMO-LUMO gap which are associated with the changes in the fluorescence emission measured in solution. Figure {\ref{Fig_STS}e} shows the associated STS spectroscopy measured for the three molecules. The FBI molecule deposited on Au(111) has an HOMO-LUMO gap of 3.17 eV. Complexation with \Nap\ slightly changes this band gap, while complexation with \Bapp\ increased it to 3.63 eV. The absence of substantial frontier molecular orbital gap upon complexation with \Nap\ is in line with other dyes containing the same crown ether, whose emission profiles do not change upon \Nap\ complexation \cite{ast_high_2011} and it is supported by the fact that no change in the molecular conformation does not alter the molecular emission. Meanwhile, the conformational change upon complexation with \Bapp\ causes a decrease in the extent of $\pi$-conjugation, which no longer extends into the phenyl ring linking the fluorophore and the crown, consequently increasing the bandgap.  The measured HOMO-LUMO gap values are in agreement with the $\lambda^{\mathrm{emi}}_{\mathrm{max}}$ spectra measured in solution, as it is summarized in Table \ref{tab:bandgaps}. The three molecules present similar $\lambda^{\mathrm{abs}}_{\mathrm{max}}$ associated to the transition of electrons from HOMO to the LUMO+n states, while the fluorescence emmision, associated to the LUMO to HOMO transition is comparable for non-chelated and \Nap-chelated molecules and shifted by 0.36 eV photon energy (489$\rightarrow$428 nm) for \Bapp\ chelated molecules. Unfortunately, because of the metallic character of the substrate we can not measure the fluorescence emission (the excitation energy is dumped to the metal without further emission). 

\section{Chelation tested on different surface support}

\begin{figure*}[ht!]
	\includegraphics[width=0.95\textwidth]{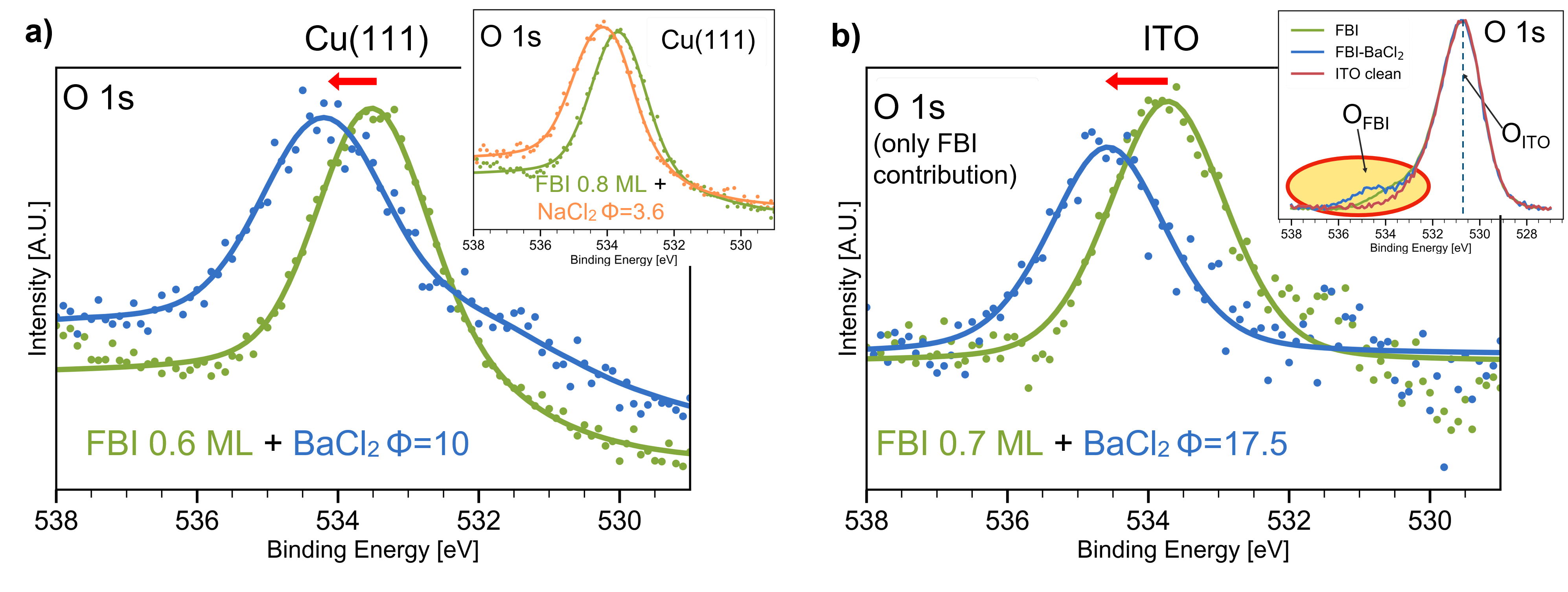}
	\caption{\label{XPS_FBI_Cu_ITO} 
    O 1s core level evolution upon chelation measured on two submonolayer FBI-functionalized surfaces: (a) chelation on Cu(111) sublimating \BappCl\ and NaCl (inset); (b) chelation on ITO. In this case, the substrate component (${\mathrm{O}_{\mathrm{ITO}}}$) was subtracted to emphasize the O 1s contribution coming from FBI (${\mathrm{O}_{\mathrm{FBI}}}$). The entire spectra are shown inserted in (b). Dot spectra correspond to raw values and solid lines to fitted curves.}
\end{figure*} 

Finally, we have tested that chelation takes place also on other surfaces, in particular Cu(111) and Indium Tin Oxide (ITO). We chose Cu(111) because it is a more reactive substrate. This enabled us to study whether the molecule-substrate interaction could alter the chelation response. On the other hand, ITO was selected as a promising candidate for the potential implementation of a barium tagging detector on a xenon-based TPC \cite{rivilla_fluorescent_2020}. As degenerated semiconductor, ITO is transparent, which would allow the direct detection of fluorescence in transmission. Furthermore, its conductivity is high enough to guide the \Bapp\ ions towards the sensor surface facilitating their capture by FBI, but potentially low enough to avoid the fluorescence quenching that is expected at conductive surfaces.

As previously discussed, the O 1s shift is a fingerprint of chelation. Figure \ref{XPS_FBI_Cu_ITO} shows the O 1s core level of FBI on Cu(111) and ITO. In both cases, upon chelation with \Bapp\ ($\phi = 10$ for 0.6 ML of FBI on Cu(111) and $\phi = 17.5$ for 0.7 ML of FBI on ITO) we observed the chemical shift on the O 1s towards higher BE, indicating that chelation is happening. In the case of Cu(111), the shift is of about 0.7 eV, while for ITO is around 0.9 eV. As far as ion capture is concerned, FBI chelation of \Nap\  on Cu(111) was also tested and, again, O 1s shows the expected upwards shift associated with chelation (inset in Figure \ref{XPS_FBI_Cu_ITO}a). These results ensure that chelation is independent from the choice of substrate.

The exact values of the O 1s core level shift are different for the three substrates. In photoemission, the absolute core level shift values depends on many factors such as the substrate, the molecular coverage, the presence of defects or other molecules. Therefore, what is important here is that the direction of the shifts is the same in the three cases, and that the magnitude of the shift is similar. 

Notice that in the case of Cu(111), when the FBI-functionalized surface is exposed to \BappCl, the residual contamination partially oxidizes the surface. For this reason, the core level has a smaller component at lower BE, around 531 eV, associated to  Cu$_2$O \cite{zhu_surface_2013}. Moreover, because of the presence of oxygen in the ITO structure, the analysis of the O 1s core level shifts required a subtraction of the core level measured on the bare ITO to enhance the FBI and \Bapp\ chelated FBI contributions (${\mathrm{O}_{\mathrm{FBI}}}$). The result of the subtraction is shown in Figure \ref{XPS_FBI_Cu_ITO} b), and the inset gathers the original normalized spectra, including the contribution to the  O 1s coming from the ITO (${\mathrm{O}_{\mathrm{ITO}}}$).

\section{Conclusions}
The demonstration of chelation in vacuum of FBI indicator by \Bapp\ ions once submonolayers of molecules are deposited on suitable surfaces is a major step towards the development of a sensor capable of single barium tagging detection, with immediate application to a new, ultra-low background xenon-based \bbonu\ detection experiment. Chemical and conformational changes occur upon chelation, independently of the substrate where molecules were deposited. These changes are furthermore in agreement with the calculations for the behaviour of free standing molecules, thus providing considerable freedom to choose the substrate. Moreover, the measured variation in the molecular HOMO-LUMO gap is perfectly consistent with the observed bicolour property of the sensor for \Bapp\ ions compared to the absence of shift observed and predicted for other ions, such as \Nap.  

Furthermore, this study has also important implications beyond neutrino particle physics field. The capability of {aza-crown ethers} to interact with many different ions is very important in applications such as drug carriers \cite{uchegbu_non-ionic_1998} or photo-switching devices \cite{malval_photoswitching_2002,uda_membrane_2005}.

\section{Methods}
The experiments were performed in two different UHV chambers, one for XPS experiments and the other for STM experiments. Both chambers have a base pressure of $1\times10^{-10}$ mbar. Prior to deposition, the Au(111) and Cu(111) and Indium Tin Oxide (ITO) surfaces were cleaned via cycles of Ar sputtering and annealing to 500 °C and their cleanliness was checked by XPS prior to molecular deposition. 

\textbf{Molecular evaporation}
Pure FBI, \BappCl\ and NaCl powders were evaporated from homemade Knudsen cell evaporators. To avoid cross-talking between FBI and the salts molecules and to exclude any possibility of chelation of the molecule inside the cell, the FBI and \BappCl\ (NaCl) evaporators were located in two separated parts of the UHV chambers. To ensure  sublimation reproducibility, the same molecular evaporator cells (containing the FBI, \BappCl\ and NaCl) were used in both STM and XPS chambers (removed from one chamber and installed in the other).  

FBI molecules were synthesized using the procedure described in \cite{rivilla_fluorescent_2020}, while \BappCl\ and NaCl were commercially (Sigma-Aldrich) available. The molecules were used after degassing in UHV.
The molecular evaporation rate was monitorized using a quartz microbalance and the amount of molecules on the surfaces was afterwards quantified by analyzing the relative intensity of the core level peaks for the experiments done in XPS as well as with the percenteage of covered surface in STM. 

\textbf{Adsoption and emission spectra}
 UV-vis spectra were acquired on a Shimatzu UV-2600 Spectrophotometer. Emission spectra were acquired on an Agilent Cary Eclipse Fluorescence Spectrophotometer. Excitation and emission monochromator bandwidth was fixed at 5 nm. Spectra were recorded at $5\times10^{-5}$ M solutions of FBI and equimolecular quantities of the corresponing perchlorate cation salt for the chelated species. Emission spectra were acquired using 250 nm light.

\textbf{STM experiments}
STM experiments were performed with a commercial Scienta-Omicron ultra-high vacuum LT-STM at 4.3 K. A W-tip  was used. Topography images were measured using constant current mode, while for bond resolution images constant height mode was used, Figure \ref{FIG_BRSTM} b-d. During the STM experiments, the tip was functionalized with CO for bond resolution STM by exposing the sample to low pressure (approximately $1 \times 10^{-8}$ mbar) of CO whilst the sample was held below 7 K. CO molecules were trapped by the tip from their adsorption sites via scanning over them or by applying $\sim$ 2 V bias voltage pulses. 

\textbf{XPS experiments}
The XPS measurements were carried out using a Phoibos-100 electron analyzer (SPECS GmbH), using a non monochromatic Al K$\alpha$ photon source of 1486.6 eV. The spectra were calibrated to the substrate main core level (Au 4f, Cu 2p, and In 3d respectively). 

The evaporation thicknesses were estimated using the attenuation of the most intense substrate core levels, i.e. Au 4f, Cu 2p and In 3d for Au(111), Cu(111) and ITO respectively. The calculations followed the guidelines provided in ref. \cite{powell_practical_2020}. For this purpose, we estimated the electron Effective Attenuation Length (EAL) through the FBI layers for electrons with kinetic energies of 1402.6, 1041.6 and 554.61 eV (Au 4f 5/2, In 3d 5/2 and Cu 2p 3/2). The resulting EALs are 3.87, 3.05 and 1.85 nm respectively. The thickness of the FBI samples was then estimated using the intensity of clean substrate core level as reference and the attenuated intensity after the evaporation. To correct for systematic uncertainties affecting different sets of data, the intensities of all spectra were rescaled to the main substrate core levels (Au 4f, Cu 2p and In 3d respectively). The core level peak area were numerically integrated between fixed local minima. This yields an error in the stoichiometry ratios of 7\%. 

The stoichiometry of the FBI (\BappCl) was calculating following the expression $ R(A/B)=(I_{A}/I_{B} \cdot S_B/S_{A})$ where A and B are the different molecular elements, $I_{X}$ is the integrated area under the main core level and $S_{X}$ is the corresponding sensitivity factor of this core level.

The amount of \Bapp (\Nap) per FBI molecule were estimated by computing  the ratio $\phi=I_{Ba}/(I_N/3) \cdot S_N/S_{Ba} $, where $I_{Ba}$, $I_N$ are total areas  of Ba 3d 5/2 and  N 1s core  levels respectively, and $S_{Ba} = 25.84$, $S_N = 1.80$ are the corresponding atomic sensitivity factors \cite{scofield_hartree-slater_1976}. The factor 3 responds to stoichiometric reasons, considering 3 N atoms per molecule.

\textbf{XPS fitting}
The spectra fitting was performed using custom-made software written in Python, using the lmfit library \footnote{https://lmfit.github.io/lmfit-py/}. An example of fit can be seen in Figure \ref{XPS_fits}. The core levels were modeled as a pseudo-Voigt function (Lorentzian to Gaussian ratio of 0.6) convoluted with a Shirley type background. The error in the estimation of the position of the maxima derived from fitting is 50 meV. However, we only consider as significant shifts higher than 100 meV, which is the energy step we use in the adquisition. The goodness of fit is given by a $\chi^2$ parameter of about 0.05 counts (in A.U.) when the data are normalized to a maximum intensity of 1. Here $\chi^2 = \sum_i (y_i- Y_i)^2/Y_i$, where $y_i$ and $Y_i$ are the data intensity values and the expected values from the fit respectively.

\begin{center}
	\includegraphics[width=0.45\textwidth]{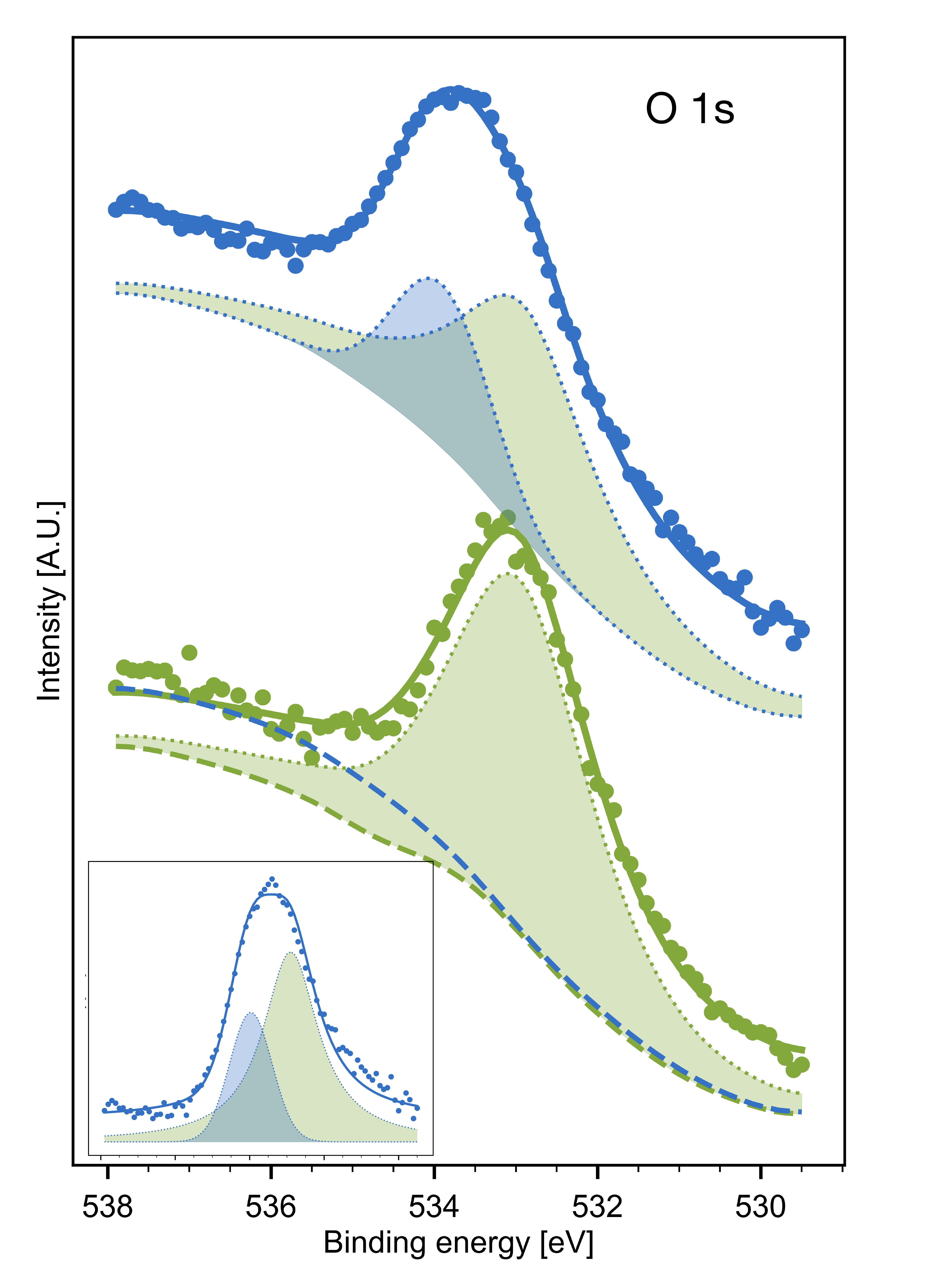}
	\captionsetup{type=figure} 
	\caption{\label{XPS_fits} Detail of fitting components (dotted lines) for O 1s in FBI 0.6 ML (green) and FBI 0.6 ML + BaCl$_2$ $\phi$ = 0.8 (blue) from Figure \ref{XPS_FBI_Au} (a) and (b). Circles and solid lines represent raw data and best fit respectively. The green component was fitted to the unchelated FBI data (green circles) at 532.95 $\pm$ 0.05 eV, with width 2.16 $\pm$ 0.04 eV at FWHM. This component was then fixed in position and width for the chelated FBI data (blue dots). An additional component was fitted to the chelated FBI CL, shown as blue filled area, at 533.91 $\pm$ 0.05 eV,  with 1.41 $\pm$ 0.05. The background of each curve is shown as dashed lines in green and blue respectively. Inset: same blue curve and components with its background subtracted.}
\end{center}

\section{Acknowledgment}
This material is based upon work supported by the following agencies and institutions: the European Research Council (ERC) under ERC-2020-SyG 951281; the MCIN/AEI/10.13039/501100011033 of Spain and ERDF A way of making Europe under grants PID2020-114252GB-I00, PID2019-107338RB-C63, PID2019-104772GB-I00, PID2019-111281GB-I00 and RTI2018-095979, the Severo Ochoa Program grant CEX2018-000867-S; the Basque Government (GV/EJ) under grants IT-1346-19, IT-1255-19, and IT-1180-19. 

The NEXT Collaboration also acknowledges support from the following agencies and institutions: the European Union's Framework Programme for Research and Innovation Horizon 2020 (2014--2020) under Grant Agreement No.\ 957202-HIDDEN; the MCIN/AEI/10.13039/501100011033 of Spain under the Mar\'ia de Maeztu Program grant MDM-2016-0692; the Generalitat Valenciana of Spain under grants PROMETEO/2021/087 and CIDEGENT/2019/049; the Portuguese FCT under project UID/FIS/04559/2020 to fund the activities of LIBPhys-UC; the Pazy Foundation (Israel) under grants 877040 and 877041; the US Department of Energy under contracts number DE-AC02-06CH11357 (Argonne National Laboratory), DE-AC02-07CH11359 (Fermi National Accelerator Laboratory), DE-FG02-13ER42020 (Texas A\&M), DE-SC0019054 (Texas Arlington) and DE-SC0019223 (Texas Arlington); the US National Science Foundation under award number NSF CHE 2004111; the Robert A. Welch Foundation under award number Y-2031-20200401. DGD acknowledges support from the Ram\'on y Cajal program (Spain) under contract number RYC-2015-18820. Finally, we are grateful to the Laboratorio Subterr\'aneo de Canfranc for hosting and supporting the NEXT experiment.

\section{Author contributions}
J.J.G.-C., F.P.C. and C.R. conceived the project. C.R. designed and coordinated the experiments and data analysis. P.H.G. and M.I. performed and analysed the XPS experiments. J.P.C., A.B.L., T.W., D.G.O. and M.C. performed the STM experiments. I.R., B.A., A.A. and Z.F. carried out the chemical synthesis. R.G.M. I.R., B.A., A.A. and Z.F. performed the characterisation in solution of fluorescence studies. P.H.G., J.P.C., J.J.G.-C., F.P.C., and C.R. wrote the manuscript. F.R. contributed to the revision of the manuscript. The NEXT collaboration assisted with editing and revision of the manuscript. 


\bibliographystyle{apsrev4-1} 
\bibliography{literature_FBI}
\end{document}